\documentclass[%
 reprint,
superscriptaddress,
nofootinbib,
 amsmath,amssymb,
 aps,
]{revtex4-2}

\usepackage{graphicx}
\usepackage{dcolumn}
\usepackage{bm}
\usepackage[version=4]{mhchem} 
\usepackage{subfig}
\usepackage{float}
\usepackage{multirow}
\usepackage[colorlinks,
            linkcolor=red,
            anchorcolor=blue,
            citecolor=blue
           ]{hyperref}

\graphicspath{ {figures/} }

\begin{document}

\title{Discrete treatment of inverse Compton scattering: implications on parameter estimation in gamma-ray astronomy}

\author{Junji Xia}
\affiliation{The School of Physical Science and Technology, Southwest Jiaotong University, Chengdu, 611756, China}
\author{Xingjian Lv}
\affiliation{Key Laboratory of Particle Astrophysics, Institute of High Energy Physics, Chinese Academy of Sciences, Beijing 100049, China}
\affiliation{University of Chinese Academy of Sciences, Beijing 100049, China}
\author{Kun Fang}
\email{fangkun@ihep.ac.cn}
\affiliation{Key Laboratory of Particle Astrophysics, Institute of High Energy Physics, Chinese Academy of Sciences, Beijing 100049, China}
\author{Siming Liu}
\email{liusm@swjtu.edu.cn}
\affiliation{The School of Physical Science and Technology, Southwest Jiaotong University, Chengdu, 611756, China}
\affiliation{Tianfu Cosmic Ray Research Center, 610000 Chengdu, Sichuan, China}



\date{\today}

\begin{abstract}
In gamma-ray astronomy and cosmic-ray physics, the continuous approximation of inverse Compton scattering (ICS) is widely adopted to model the evolution of electron energy. However, when the initial electron energy approaches $\sim100$ TeV, the discrete nature of ICS becomes prominent, and the energy of evolved electrons should be considered as a broad distribution rather than a deterministic value. By simulating the evolution paths of individual electrons under ICS, we capture this discrete nature and demonstrate that when the electron injection spectrum exhibits a high-energy cutoff, the correct discrete treatment yields a higher cutoff energy in the evolved spectrum compared to the continuous approximation. Applying the discrete ICS treatment to interpret the gamma-ray spectrum of the Geminga pulsar halo measured by HAWC, we find that the inferred cutoff energy of the injection spectrum is correspondingly lower than that derived using the continuous approximation at a $95\%$ confidence level. This suggests that the systematic bias introduced by the approximation has exceeded the measurement precision. We also expect the application of the discrete ICS correction in the PeV regime using the ultra-high-energy gamma-ray source 1LHAASO J1954+2836u as a case study, pointing out that adopting the continuous approximation may considerably overestimate the electron acceleration capability of the source.
\end{abstract}
\maketitle


\section{Introduction}
\label{sec:intro}
Inverse Compton scattering (ICS) of relativistic electrons with low-energy photons is a fundamental radiation mechanism in gamma-ray astronomy. ICS also plays a dominant role in determining the energy evolution of the parent electrons of gamma rays. Unlike the synchrotron radiation process of electrons in magnetic fields, ICS between electrons and background photons is a discrete process. In practical calculations, the energy evolution of electrons is generally approximated using a continuous treatment, represented by the equation $dE_e/dt = -b_0(E_e)E_e^2$, where $E_e$ is the electron energy, and $b_0$ is an energy-dependent coefficient that accounts for the Klein-Nishina effect (e.g., see \cite{Fang:2020dmi}). However, as the electron energy increases, the frequency of ICS decreases, and the energy loss per scattering becomes significant compared to the initial electron energy. For example, a $100$~TeV electron can lose about half its energy in a single scattering with a microwave background (CMB) photon \cite{John:2022asa}, where the continuous approximation of ICS clearly breaks down. This issue had already been noted by the classic paper on radiation mechanisms in astrophysics \cite{Blumenthal:1970gc}; however, due to the observational limitations at the time, rigorous calculations did not yet have a clear application scenario.

Gamma-ray astronomy has entered the ultra-high-energy (UHE, $>100$~TeV) era, with the energy spectra of over 30 gamma-ray sources measured beyond $100$~TeV \cite{LHAASO:2023rpg}. Among them, about 20 UHE gamma-ray sources are associated with pulsars. These sources are likely related to pulsar wind nebulae (PWNe) or pulsar halos and may thus originate from ICS. Apparently, when constructing evolution models for the parent electrons of these sources, the impact of the discrete nature of ICS must be discussed. Owing to the randomness of discrete scatterings, the evolved energy of electrons is no longer a deterministic value but instead forms a distribution with a considerable width. This broadening effect directly influences the calculated gamma-ray energy spectrum. As a result, when we deduce the initial electron injection spectrum from the observed gamma-ray spectrum using a continuous approximation of ICS, the inferred electron injection spectrum will appear broader than the true one. Therefore, when the gamma-ray spectrum exhibits structures deviating from a power law, a discrete treatment of ICS would be necessary.

In this work, we investigate the impact of the discrete treatment of ICS on the parameter estimation of electron injection spectra in gamma-ray astronomy. We use a Monte Carlo method to simulate the energy evolution of electrons due to ICS. This approach is similar to those adopted in refs.~\cite{John:2022asa} and \cite{John:2023ulx}, which investigate the necessity of discrete treatment of ICS in dark matter indirect detection. In Section~\ref{sec:single}, we present the simulation method for the energy evolution of individual electrons due to ICS. In Section~\ref{sec:electron}, we apply the discrete treatment of ICS to time-dependent electron injection spectra. In Section~\ref{sec:gamma}, by fitting the spectra of specific gamma-ray sources, we demonstrate the significant differences in the inferred electron injection spectra between the discrete treatment and continuous approximation. Section~\ref{sec:conclu} provides the conclusion of the paper.

\section{Discrete ICS for single-energy electrons}
\label{sec:single}
The energy evolution equation of electrons due to ICS is typically written as
\begin{equation}
 \frac{dE_e}{dt}=-\iint (E_\gamma-\epsilon)f(E_e, E_\gamma, \epsilon)dE_\gamma d\epsilon \,,
 \label{eq:continuous}
\end{equation}
where $f(E_e, E_\gamma, \epsilon)\equiv dN/(dtdE_\gamma d\epsilon)$ represents the scattering rate at which a electron scatters with a background photon of energy $\epsilon$ to produce a gamma photon of energy $E_\gamma$. Equation (\ref{eq:continuous}) integrates over the energies of the background photons and the emitted photons, treating the energy evolution of electrons as a smooth process over time. This is referred to as the continuous approximation of ICS, which is useful for estimating the \textit{average} energy evolution behavior of a large number of electrons. However, for individual electrons, ICS occurs infrequently, and the energy of emitted photons from each scattering event exhibits substantial randomness. If we consider only CMB as the background photon field, the average time interval between two scatterings for a $100$~TeV electron is estimated to be $\approx6$~kyr, and the average energy loss per collision is $\approx20$~TeV. This means that the evolution path of each electron appears as a distinct step-like pattern, which may deviate significantly from the statistical average given by Eq.~(\ref{eq:continuous}). The higher the initial energy of the electron, the larger the expected energy loss per collision is in comparison to the initial energy, and the more pronounced the discreteness\footnote{For comparison, a $1$~TeV electron has an expected energy loss of $2$~GeV per collision, which is only $0.2\%$ of its initial energy.}.

To simulate the discrete nature of the evolution path of a single electron, we adopt the Monte Carlo method. The sampling is based on the scattering rate in Eq.~(\ref{eq:continuous}), which can be expressed as follows (as given by ref.~\cite{Blumenthal:1970gc}):
\begin{equation}
 \begin{aligned}
  f(E_e, E_\gamma, \epsilon) & = \frac{3\sigma_Tc}{4(E_e / m_e c^2)^2} \times \frac{n(\epsilon)}{\epsilon} \\
  & \times \left[ 2q \ln q + (1 + 2q)(1 - q) + \frac{(\Gamma q)^2 (1 - q)}{2(1 + \Gamma q)} \right]\,,
 \end{aligned}
\label{eq:rate}
\end{equation}
where
\begin{equation}
\Gamma = \frac{4E_e\epsilon}{(m_ec^2)^2}\,, \quad
q = \frac{E_\gamma / E_e}{(1 - E_\gamma / E_e) \Gamma}\,,
\end{equation}
and $n(\epsilon)$ is the differential energy density of background photons, while \( \sigma_T \) denotes the Thomson cross-section. The photon fields we adopt include the CMB, the infrared radiation from dust, and the starlight radiation, with their temperatures and energy densities being ($2.7$~K, $0.26$~eV~cm$^{-3}$), ($20$~K, $0.3$~eV~cm$^{-3}$), and ($5000$~K, $0.3$~eV~cm$^{-3}$), respectively \cite{Abeysekara:2017old}. For the dust and starlight background, we assume that $n(\epsilon)$ follows a graybody spectrum, that is, it adopts the spectral shape of blackbody radiation but is renormalized according to its energy density.

\begin{figure}[t]
    \includegraphics[width=0.45\textwidth]{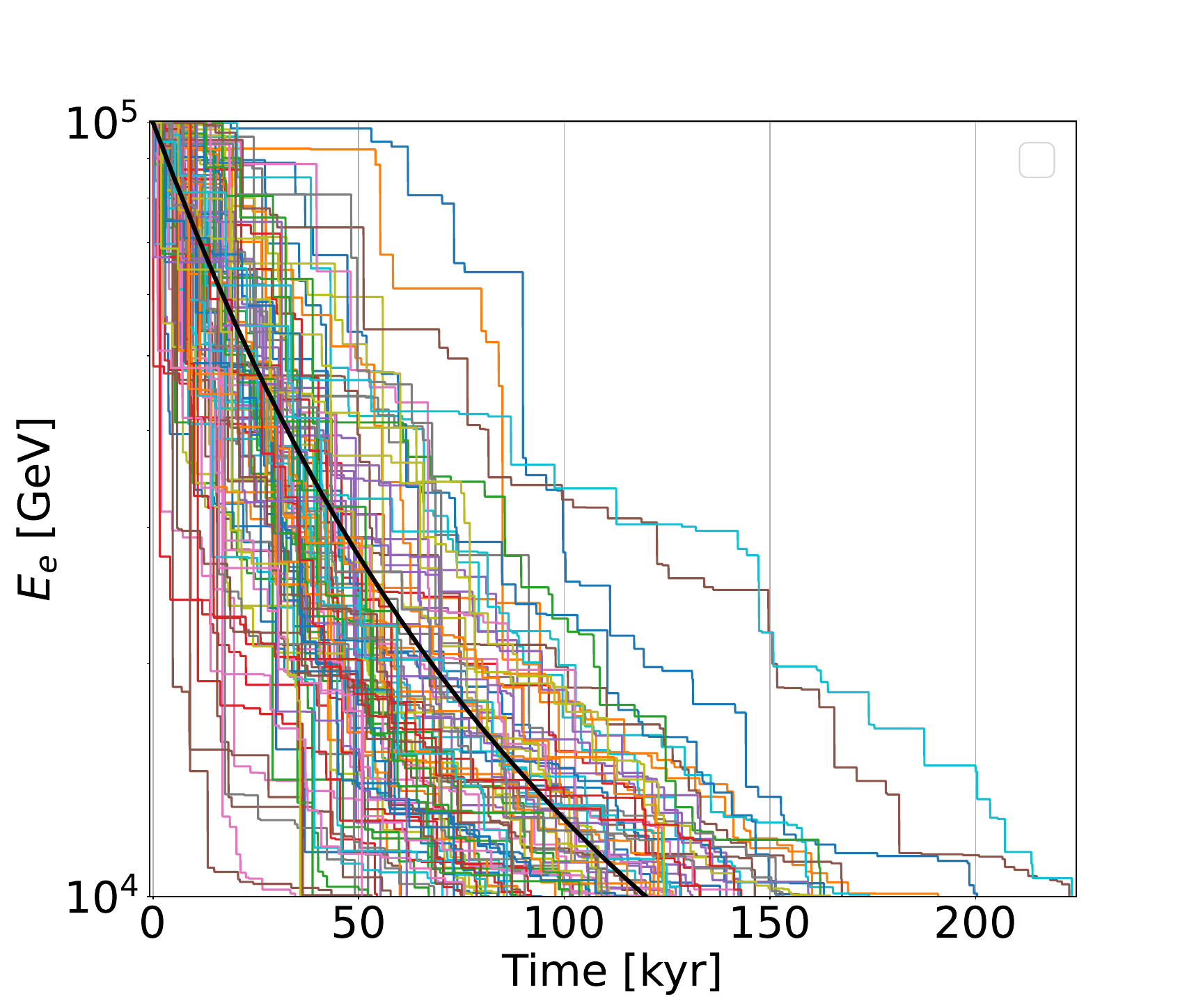}
    \caption{The colored lines illustrate the energy evolution paths of 100 electrons due to ICS, as given by the Monte Carlo simulation. All the electrons have an initial energy of 100 TeV. For comparison, the black line represents the electron energy evolution under the continuous approximation of ICS.}
    \label{fig: E_e}
\end{figure}

We choose a sampling time step of $\Delta t = 50$~yr, which is much shorter than the average time interval between two ICS events, $T=1/\iint f(E_e, E_\gamma, \epsilon)\,dE_\gamma\,d\epsilon$. This ensures that at most one scattering event can occur within each step. In each sampling time step, the probability of a scattering occurring is $\Delta t/T$, which we use to determine if scattering will happen. If a scattering event does occur, we apply Eq.~(\ref{eq:rate}) as the weight to perform random sampling for both the background photon energy $\epsilon$ and the emitted photon energy $E_\gamma$. Consequently, we obtain the electron energy loss resulting from this scattering event, i.e., $E_\gamma-\epsilon$. Based on this method, we present in Fig.~\ref{fig: E_e} the evolution paths of 100 electrons with an initial energy of $100$~TeV. It can be seen that the evolution paths of individual electrons can deviate significantly from the statistical average given by the continuous approximation.

\begin{figure*}[htbp]
    \begin{center}
        \subfloat{\includegraphics[width=0.42\textwidth]{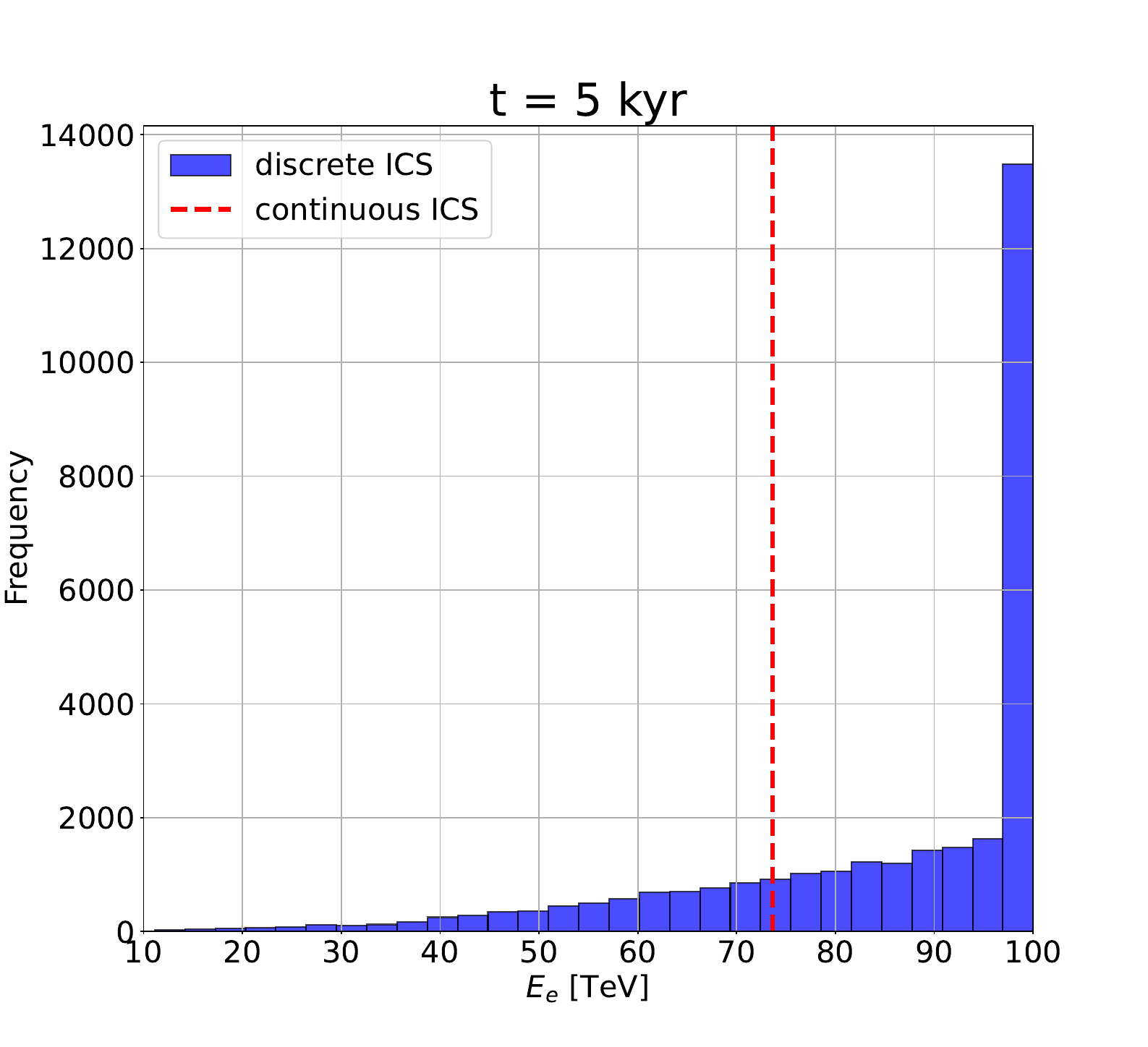}} \hskip 0.01\textwidth
        \subfloat{\includegraphics[width=0.4\textwidth]{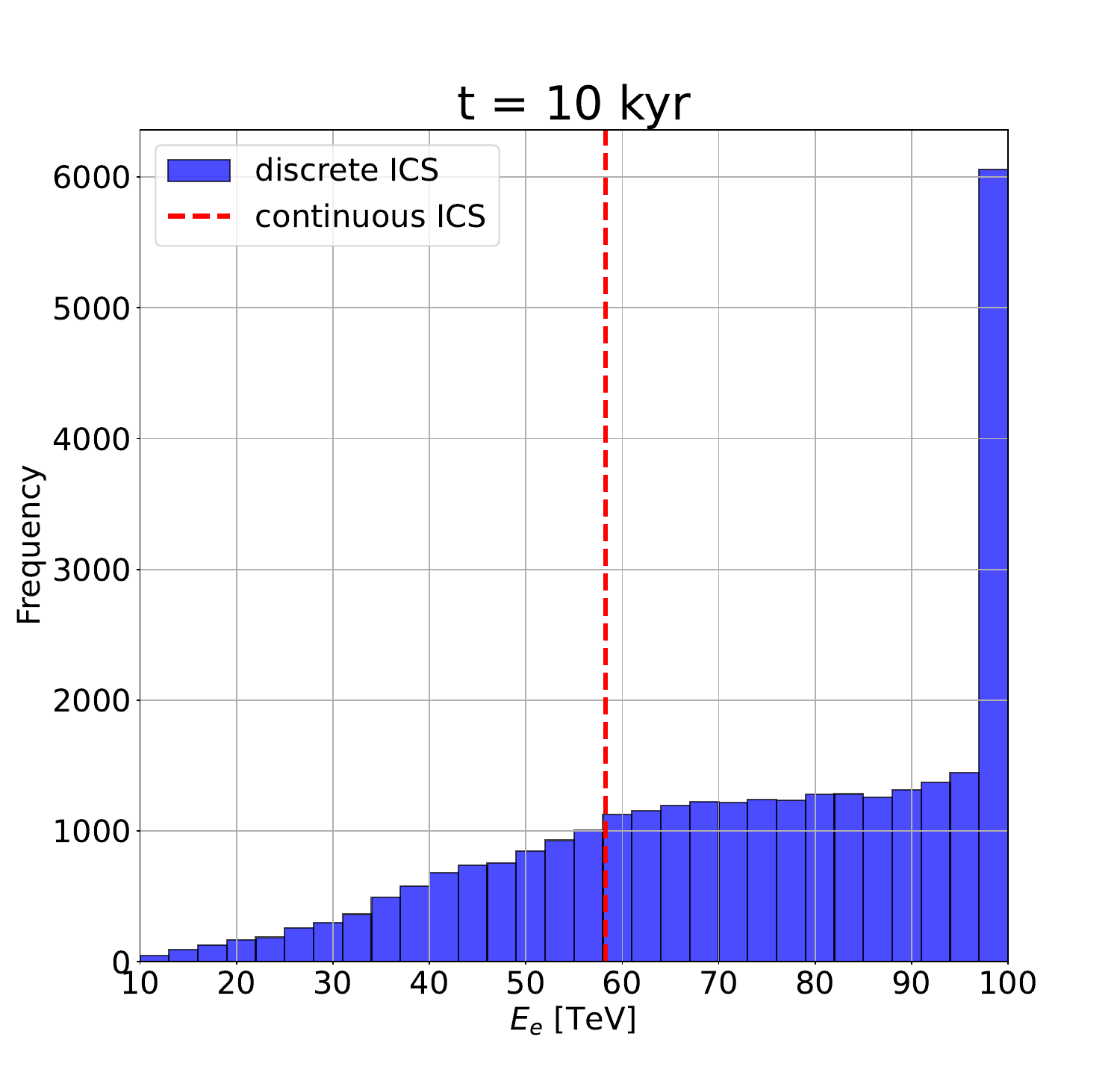}}
        \hskip 0.01\textwidth
        \subfloat{\includegraphics[width=0.4\textwidth]{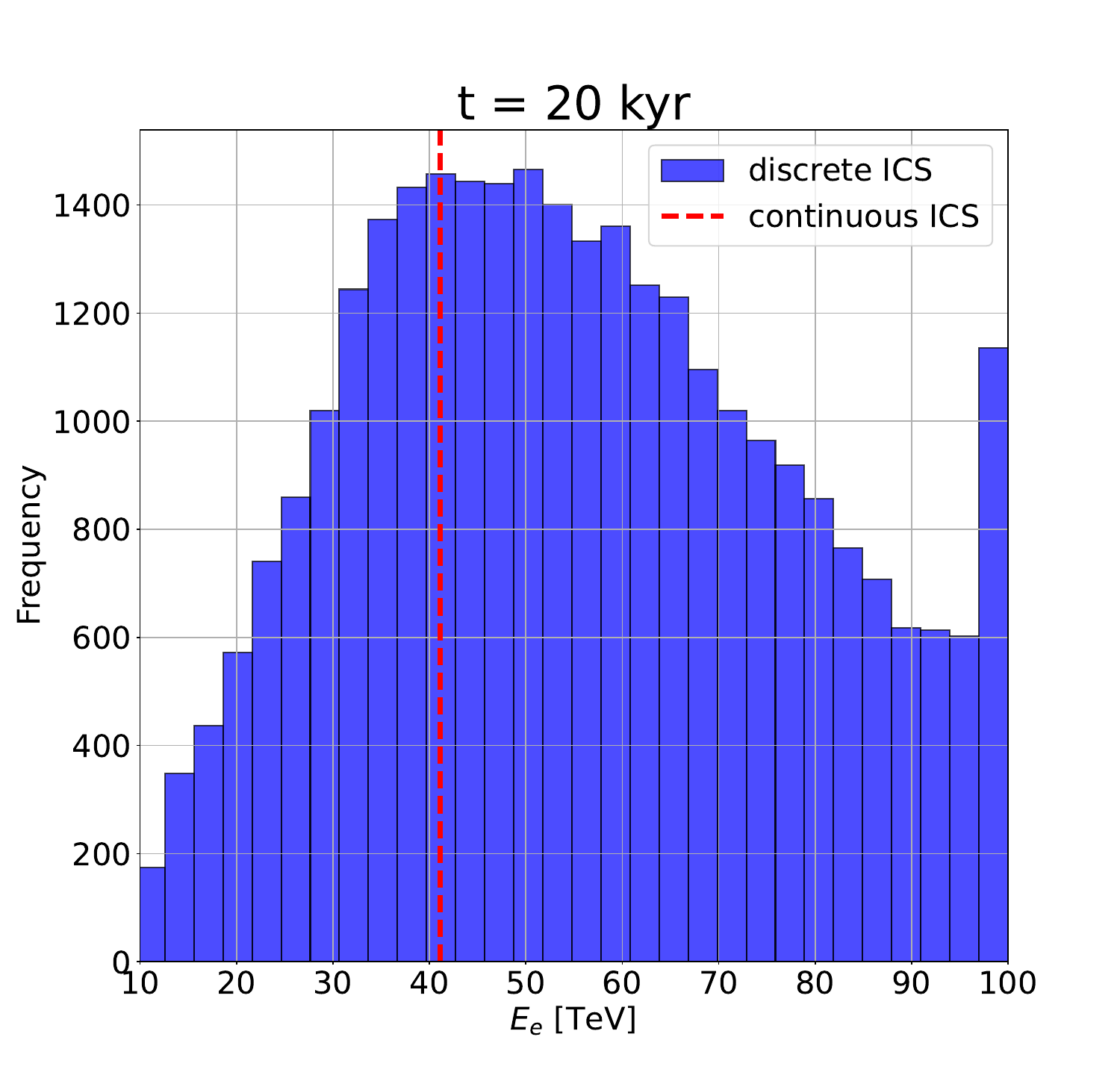}}
        \hskip 0.01\textwidth
        \subfloat{\includegraphics[width=0.4\textwidth]{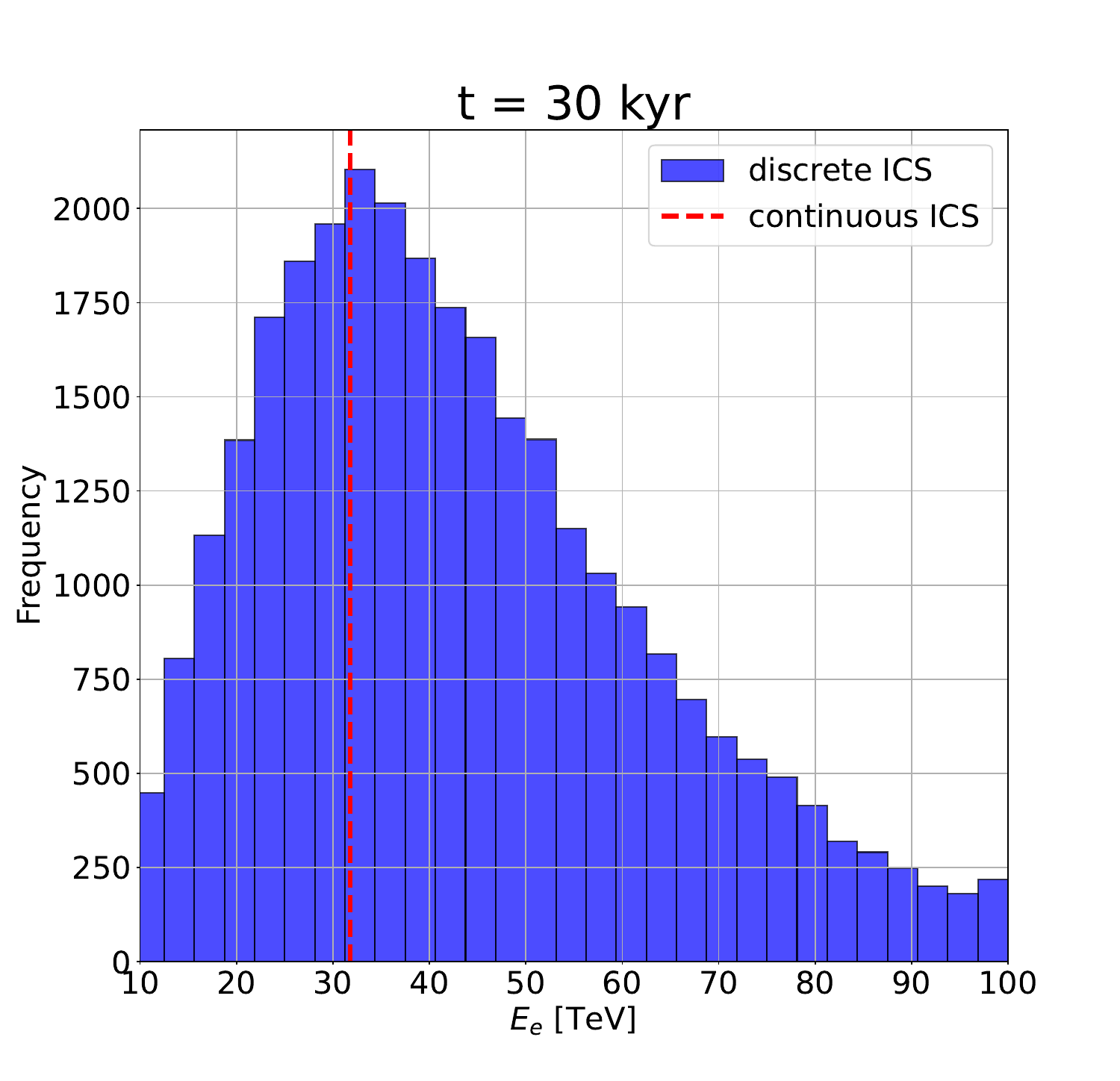}}      
    \end{center}
    \caption{Electron energy distributions at various evolution times, which are statistically obtained from the simulation of ICS involving 30,000 electrons. All electrons have an initial energy of $100$~TeV. For comparison, the red dashed line represents the electron energy expected from the continuous approximation of ICS.}
    \label{fig: distribution}
\end{figure*}

To more clearly demonstrate the energy dispersion of electrons under discrete treatment, we simulate 30,000 electrons, each with an initial energy of $100$ TeV, and statistically analyze the electron energy distribution at various evolution times, as shown in Fig.~\ref{fig: distribution}. When $t=5$~kyr, the evolution time is shorter than the scattering period of electrons with $E_e=100$~TeV, so a large number of electrons have not yet undergone ICS, and the energy distribution of electrons differs significantly from the continuous approximation expectation. 

As evolution time increases, the proportion of electrons that have not reacted decreases, and the electron energy distribution shifts towards a Gaussian-like shape. The center of the distribution tends to be consistent with the energy predicted by the continuous approximation; however, the width of the distribution is quite significant compared to the central energy. Moreover, the distribution width does not decrease over time. The reason is that at the early stages of evolution, ICS of high-energy electrons exhibit significant discrete effects, and the evolution paths of individual electrons diverge noticeably. However, as the energy decreases, the discrete effects begin to weaken, and the fluctuations in the evolution paths become smaller. Consequently, the deviations brought about by the early evolution of electrons are preserved. This indicates that the discrete treatment of ICS remains significant for relatively old electron sources.

\section{Corrected parent electron spectrum}
\label{sec:electron}

\begin{figure*}[htbp]
    \begin{center}
        \subfloat{\includegraphics[width=0.47\textwidth]{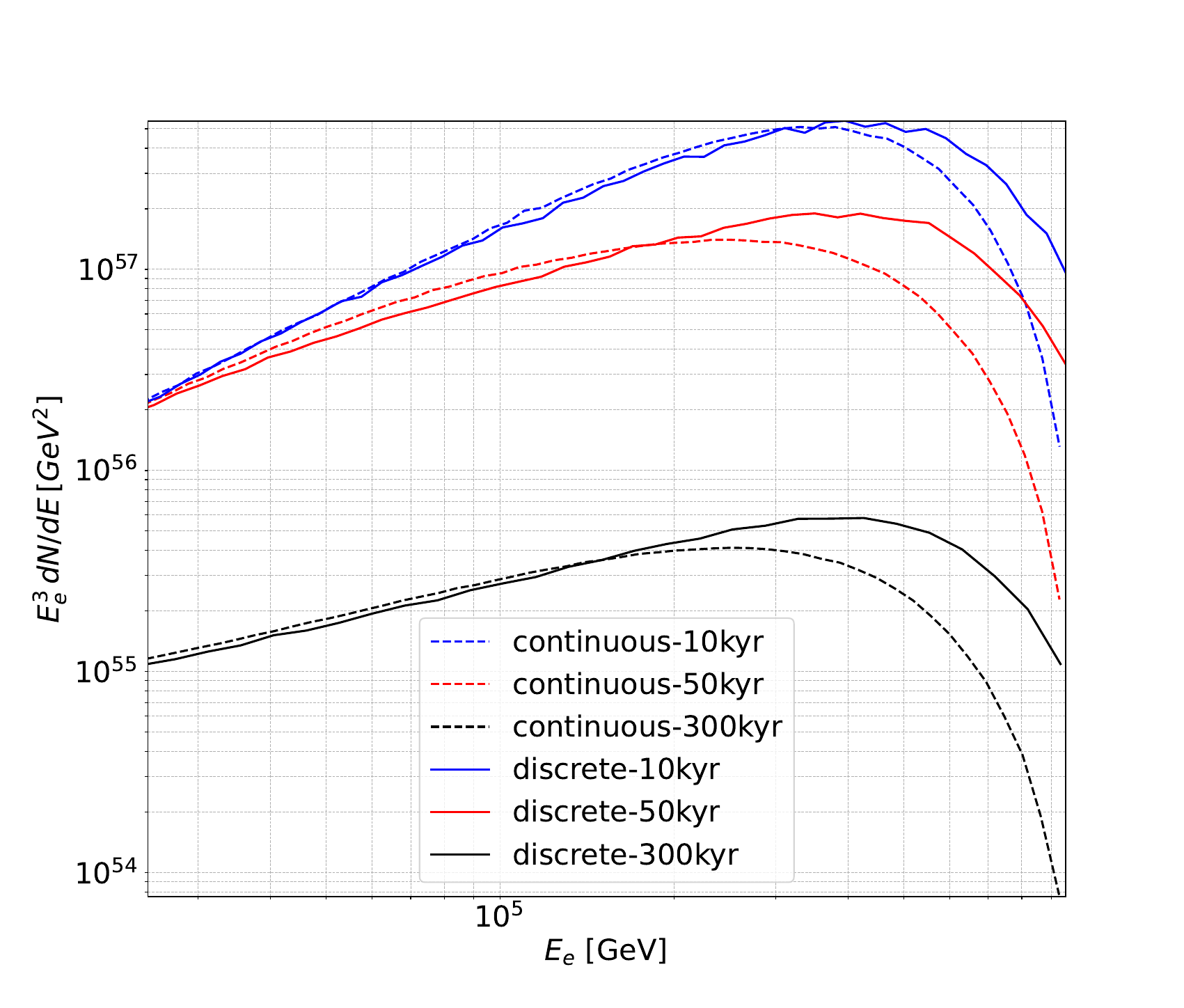}}
        \subfloat{\includegraphics[width=0.47\textwidth]{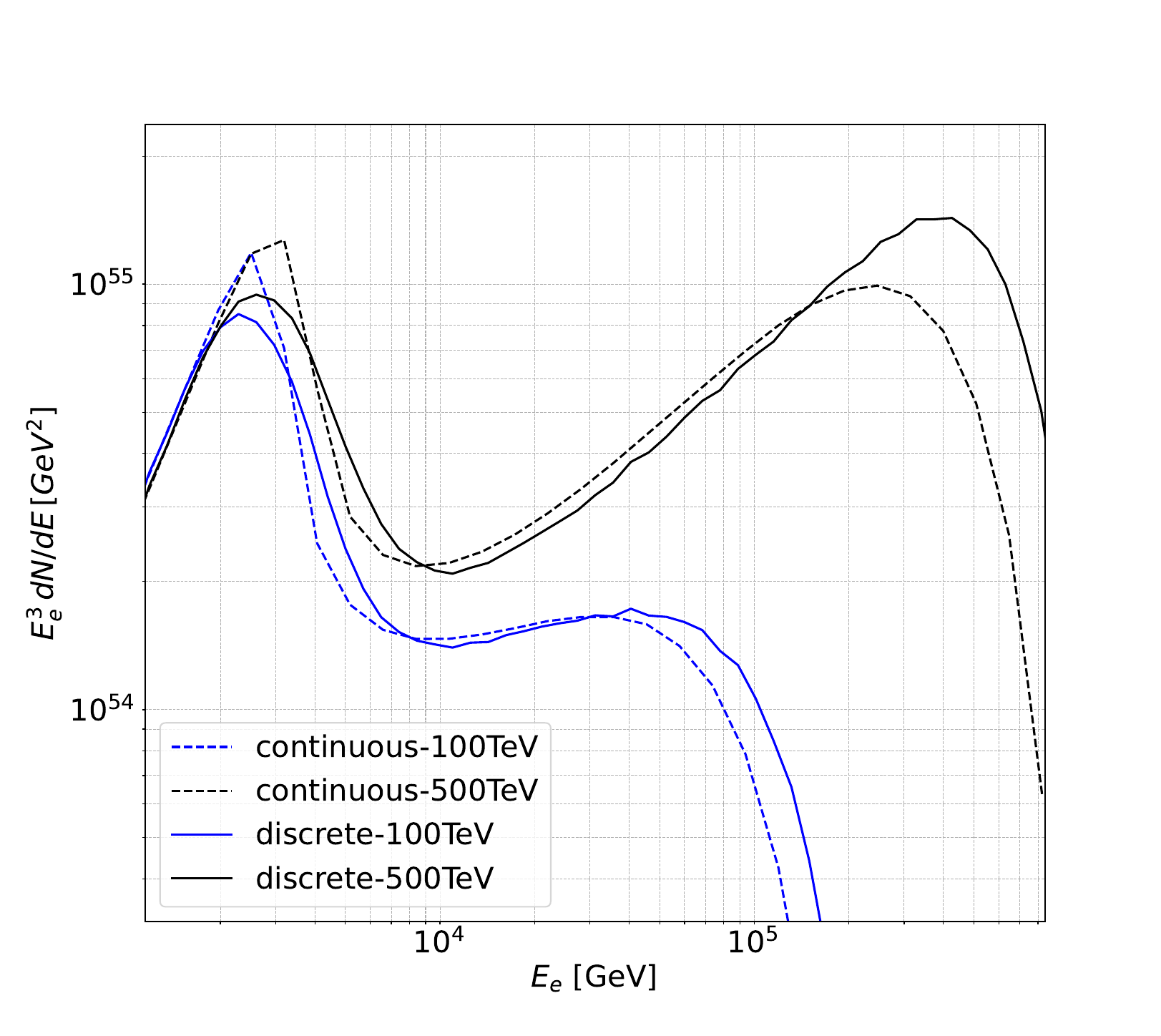}} \hskip 0.01\textwidth
    \end{center}
    \caption{Impact of the discrete ICS correction on the evolved electron spectrum, assuming a PWN as the electron source. The left panel is for various evolution times of the pulsar, while the right is for different cutoff energies of the electron injetion spectrum.}
    \label{fig: diff}
\end{figure*}

In order to discuss the significance of the discrete correction of ICS in gamma-ray astronomy, we extend the simulation in Section \ref{sec:single} to the case of a time-dependent electron injection spectrum. Among high-energy astrophysical sources, PWNe can accelerate electrons with $E_e>100$~TeV over extended periods through the continuous generation of relativistic shocks \cite{Gaensler:2006ua,Reynolds:2017hbs}. Therefore, PWNe and pulsar halos, which result from the escaped electrons of PWNe \cite{Fang:2022fof,Liu:2022hqf}, provide suitable scenarios for applying the discrete correction of ICS. We construct models based on the characteristics of these sources, as outlined below.

The electron injection spectrum is assumed to be
\begin{equation}
 \frac{dN(E_e,t)}{dE_e} = Q(t)\,E_e^{-\alpha} e^{-(E_e/E_{e, \mathrm{cut}})^2}\,,
 \label{eq:inj}
\end{equation}
where the energy dependence is suggested by the relativistic acceleration theory \cite{Dempsey:2007ng}. The time profile is assumed to follow the temporal evolution of the spin-down luminosity of pulsars, that is,
\begin{equation}
 Q(t) \propto \left( 1 + \frac{t}{\tau} \right)^{-2},
 \label{eq:qt}
\end{equation}
where $\tau$ is the spin-down timescale and is set to be $10$~kyr.

Our injected electron energy distribution ranges from 1 TeV to 1 PeV. Different from the approach in ref.~\cite{John:2022asa}, we do not determine the initial number of samples in each energy bin based on the spectrum in Eq.~(\ref{eq:inj}). Instead, we assign 100 electrons to each energy bin. This approach ensures an adequate number of samples in each bin, preventing an excess of low-energy samples and a shortage of high-energy samples. We inject electrons starting from $t = 0$ at intervals of $1$~kyr until $t$ reaches the current pulsar age. After all electrons have evolved, we renormalize the number of electrons according to their initial energy and injection time using the energy and time dependence in Eq.~(\ref{eq:inj}). Finally, we perform time integration to obtain the evolved energy spectrum.

On the left panel of Fig.~\ref{fig: diff}, we compare the electron energy spectra obtained by the discrete ICS method with those expected from the continuous approximation at different evolution times. The injection spectrum parameters are both assumed to be $\alpha = 1.5$ and $E_{e,\mathrm{cut}} = 500$~TeV. The difference between the discrete method and the continuous approximation is mainly in the high-energy cutoff region. Given the same injection spectrum, the cutoff energy of the evolved spectrum obtained by the continuous approximation is lower. Physically, this is because the correct discrete treatment, compared to the continuous approximation, allows some high-energy electrons to have a longer lifetime. As can be seen from Fig.~\ref{fig: distribution}, the discrete correction of ICS is approximately equivalent to convolving the spectrum of the continuous approximation with an energy spread function. If the injection spectrum is a pure power-law spectrum, this does not lead to significant changes. However, in cases where the spectrum includes a cutoff, the relative scarcity of high-energy electrons makes the spread of low-energy electrons to higher energies—due to the discrete correction—substantially increase the number of high-energy electrons. 

From the left panel of Fig.~\ref{fig: diff}, we find that the impact of the correction due to discrete treatment grows with time and begins to stabilize at $t = 50$~kyr (the correction magnitude at $t = 300$~kyr shows no obvious difference from that at $t = 50$~kyr). The reason is that when $t \geq 50$~kyr, the high-energy part of the electron energy spectrum is always dominated by recently injected components. Therefore, for PWNe or pulsar halos associated with middle-aged pulsars, it remains important to consider the discrete correction of ICS.

The right panel of Fig.~\ref{fig: diff} presents the results for different $E_{e,\mathrm{cut}}$ at the same evolution time ($t=300$~kyr). The effect of the discrete correction is more significant for a higher cutoff energy of the injection spectrum. As discussed in Section~\ref{sec:single}, electrons with higher energies suffer greater fluctuations in their energy-loss path. This figure additionally illustrates the effect of the discrete correction of ICS at multi-TeV energies. The sharp spectral structure, which arises from the synergy of the time evolution of the injection rate and the energy loss of electrons, is smoothed due to the discrete correction. This is consistent with the findings of ref.~\cite{John:2022asa}. 

These results indicate that when the evolved electron spectrum deviates significantly from a power law, such as exhibiting a spectral cutoff or bump, the discrete correction of ICS becomes essential.

\section{Impact on parameter estimation}
\label{sec:gamma}
Section~\ref{sec:electron} illustrates that if the continuous approximation of ICS is used to calculate the electron energy loss, the cutoff energy of the obtained electron spectrum will be lower than the correct value. In other words, when we interpret the gamma-ray spectrum emitted by electrons, the continuous approximation can lead to an overestimation of the inferred cutoff energy of the injection spectrum. In this section, we explore the importance of the discrete treatment of ICS for parameter estimation in gamma-ray astronomy, using two specific high-energy gamma-ray sources, the Geminga halo and 1LHAASO J1954$+$2836u, as examples.

\subsection{The Geminga halo}
\label{subsec:geminga}
The Geminga halo is the prototype of gamma-ray pulsar halos \cite{Abeysekara:2017old,HESS:2023sbf}. It is generated by ICS of high-energy electrons accelerated and released by the Geminga PWN, and the slow-diffusion phenomenon near the source ensures the high brightness of the Geminga halo. Although Geminga is a middle-aged pulsar, x-ray observations indicate that its PWN still possesses the ability to accelerate electrons to $\sim100$~TeV \cite{2003Sci...301.1345C,Posselt:2016lot}. Currently, HAWC has conducted precise measurements of the gamma-ray spectrum of the Geminga halo over an energy range covering two orders of magnitude \cite{HAWC:2024scl}. Compared to x-ray observations, this offers a superior opportunity to determine the electron injection spectrum of the Geminga PWN, which is crucial for understanding its acceleration capability.

In this work, we do not consider the spatial transport processes and only focus on the evolution of the electron spectrum due to energy losses. We fit the spectrum of the Geminga halo measured by HAWC using the continuous and discrete treatments of ICS, respectively. Observations of the region within $1^\circ$ around Geminga by eROSITA suggest that the magnetic field strength in the surrounding interstellar medium could be lower than $\approx1.5$~$\mu$G \cite{Khokhriakova:2023rqh}. Therefore, when calculating the total energy loss, we assume $B=1$~$\mu$G for the synchrotron term.

\begin{figure*}[htbp]
    \begin{center}
        \subfloat{\includegraphics[width=0.45\textwidth]{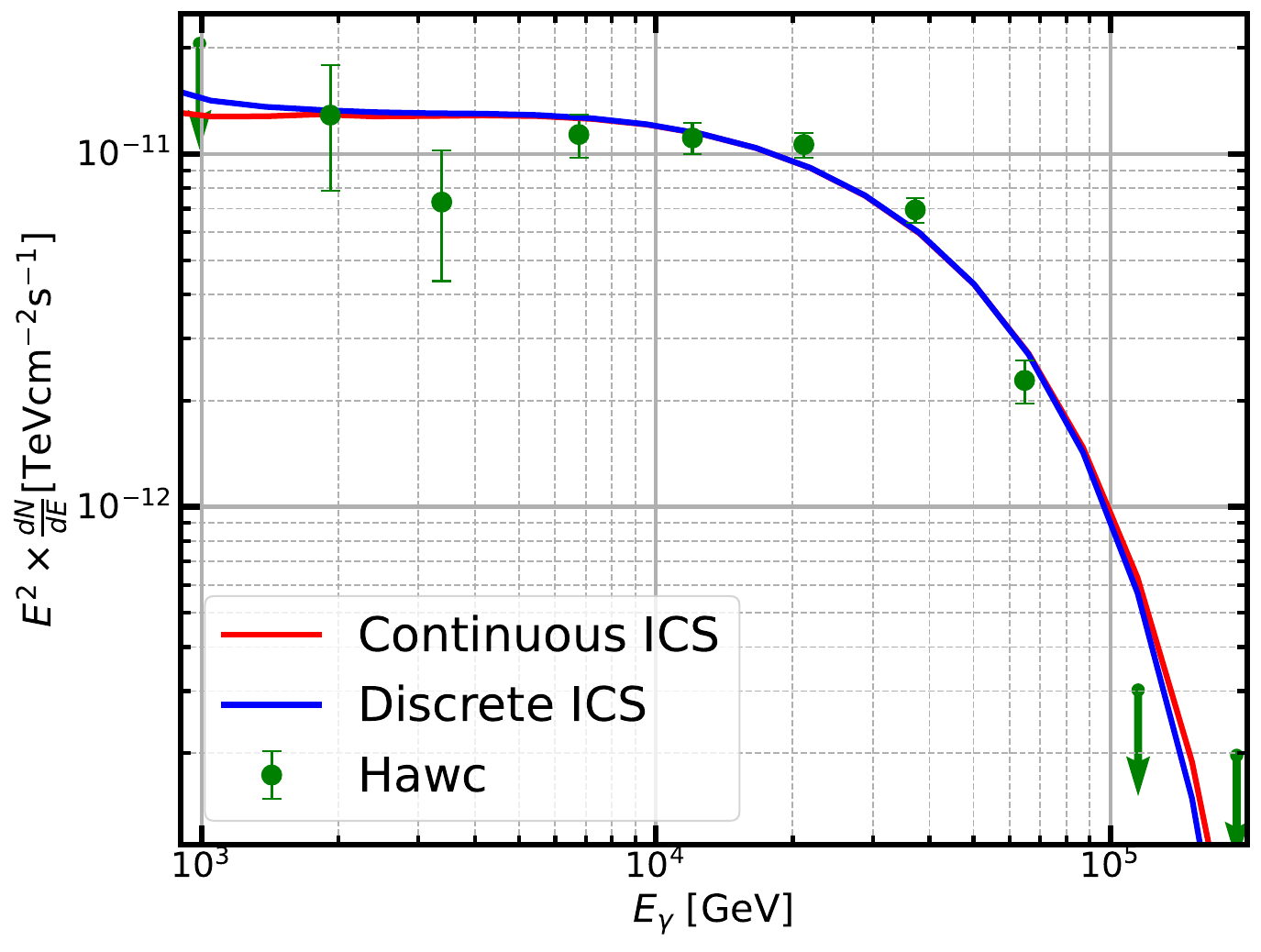}} \hskip 0.01\textwidth
        \subfloat{\includegraphics[width=0.43\textwidth]{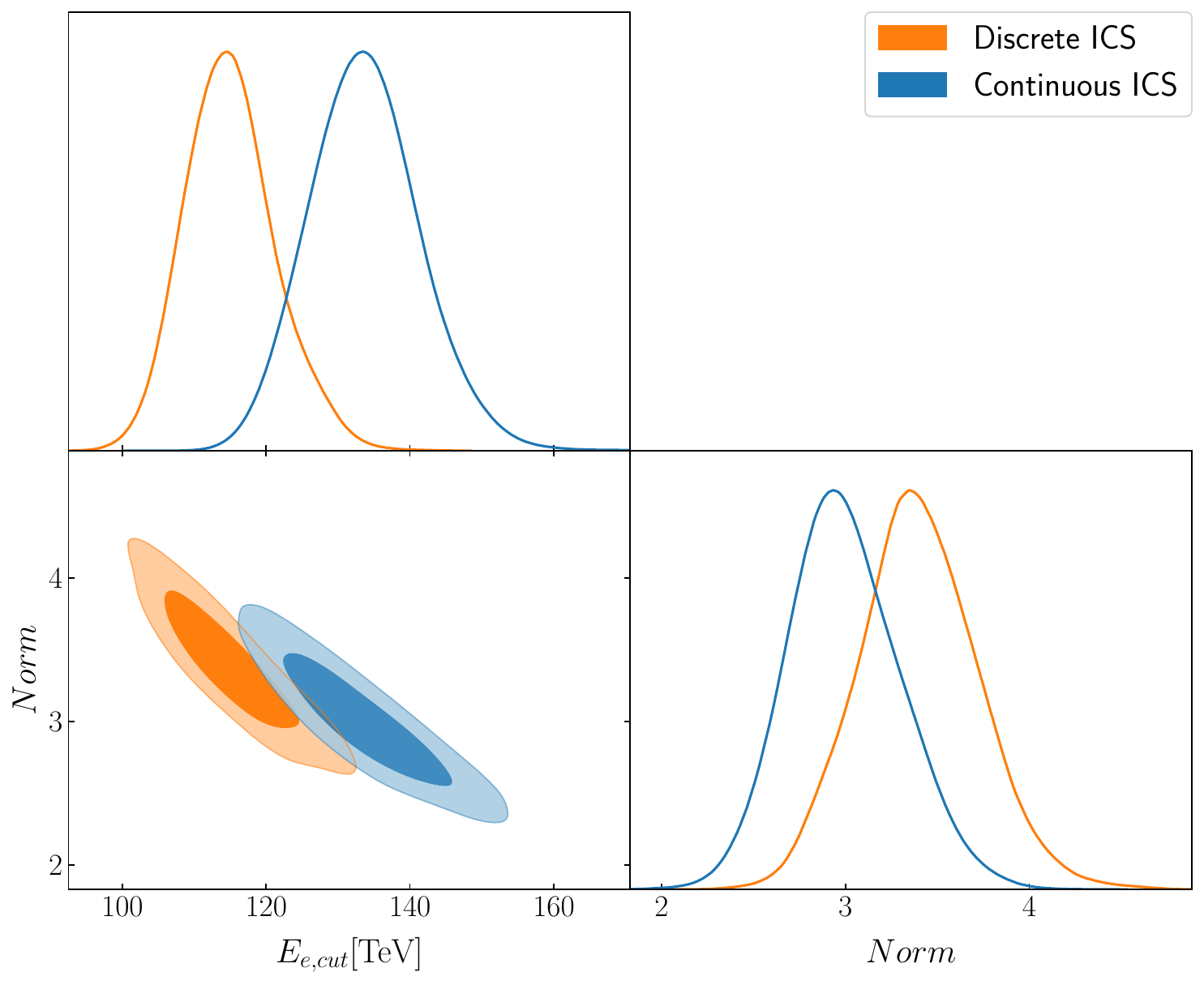}}
    \end{center}
    \caption{Left: Best-fit models for the gamma-ray spectrum of the Geminga halo measured by HAWC under the discrete treatment and continuous approximation of ICS, respectively. Right: 1D and 2D posterior distributions of fitting parameters derived from Bayesian analysis.}\label{fig: geminga}
\end{figure*}

In the fitting procedure, we consider the cutoff energy $E_{e,\mathrm{cut}}$ and a normalization factor of the spectrum as free parameters. Given that the discrete correction of ICS does not significantly influence the determination of the power-law index, we fix it as $\alpha=1.0$\footnote{If $\alpha$ is regarded as a free parameter, it will tend to values much smaller than those expected by acceleration theory or simulations in the fitting process. This could be due to the finite size of the slow-diffusion region around Geminga, indicating the gamma-ray spectrum measured by HAWC may not reflect the whole-space spectrum \cite{Fang:2023xla}. This conflict can be solved by employing a two-zone diffusion model \cite{Fang:2023xla}. However, this is beyond the scope of this work.}, which is consistent with the suggestions from x-ray observations and the expectations of simulations \cite{Posselt:2016lot,Bykov:2017xpo}. We estimate the posterior distribution of the parameters using Bayesian theory and explore the parameter space through the Markov Chain Monte Carlo (MCMC) method implemented by the Python package \texttt{Cobaya}\footnote{\url{https://cobaya.readthedocs.io/en/latest/}}~\cite{Lewis:2002ah, Lewis:2013hha, Torrado:2020dgo}. We adopt uniform priors for both parameters. The fitting results are presented in Fig.~\ref{fig: geminga} and Table~\ref{table: Geminga}.

\begin{table}[htbp]
\captionsetup{justification=raggedright}
\caption{The prior ranges, best-fit values, mean values, and posterior 68\% range of the injection parameters for the Geminga halo using the discrete and continuous ICS energy losses, respectively. $E_{e,\mathrm{cut}}$ is the cutoff energy of the injection spectrum, and Norm is the total number of electrons from $1$~TeV to $1$~PeV. \label{table: Geminga}}
\begin{tabular}{lcccccc}
\hline\hline
\multirow{2}{*}{}               & \multirow{2}{*}{Prior Range} & \multicolumn{2}{c}{Discrete ICS}                           &  & \multicolumn{2}{c}{Continuous ICS}                             \\ \cline{3-4} \cline{6-7} 
                                &                              & Best                         & Mean                          &  & Best                         & Mean                          \\ \hline
$E_{e,\mathrm{cut}}$/TeV                         & {[}50, 250{]}                & 114.9                       & $115.0^{+5.4}_{-6.9}       $  &  & 133.8                       & $133.7^{+7.1}_{-8.1}       $  \\
Norm/$10^{44}$                         & {[}0.5,10{]}                & 3.40                       & $3.40^{+0.18}_{-0.18}      $  &  & 2.97                       & $2.97^{+0.16}_{-0.19}      $  \\                         \hline
\hline
\end{tabular}
\end{table}

The left panel of Fig.~\ref{fig: geminga} shows that both ICS treatments achieve comparable fits to the data. However, the posterior distributions in the right panel reveal substantial differences in the parameter estimates derived from the two methods: the cutoff energy obtained by the discrete method is $115.0^{+5.4}_{-6.9}$~TeV, while the result of the continuous approximation is $133.7^{+7.1}_{-8.1}$~TeV, which is higher than the accurate value with a confidence level of $\approx95\%$. This indicates that the systematic bias in parameter estimation due to the inaccurate ICS calculation has exceeded the statistical error of experimental measurements. Currently, gamma-ray astronomy has entered the era of precise measurements. We argue that to attribute physical significance to the outcomes of precise measurements, it is essential to employ rigorous calculation methods.



\subsection{1LHAASO J1954$+$2836u}
\label{subsec:j1954}
Section \ref{sec:electron} illustrates that the correction in parameter estimation due to discrete ICS becomes more significant as the cutoff energy of the electron injection spectrum increases. The gamma-ray spectrum of the Geminga halo exhibits a roll-off at tens of TeV, while LHAASO has already observed a group of sources with spectra extending close to PeV. In theory, these sources are more suitable candidates for applying discrete ICS. However, since the spectra of these UHE sources are generally soft above $100$~TeV, any spectral cutoff features would result in a very low expected flux, making detection challenging at present.

With the accumulation of data, LHAASO may eventually detect the spectral cutoffs of some of the sources. Taking 1LHAASO J1954$+$2836u (2HWC J1955$+$285 \cite{Abeysekara:2017hyn}) as an example, we offer insight into the application of discrete ICS treatment. 1LHAASO J1954$+$2836u is classified as a PWN in the TeV online catalog \cite{Wakely:2007qpa}, and the characteristic age of the corresponding pulsar is $\approx70$~kyr. According to the calculations in Section~\ref{sec:electron}, the discrepancy between the results of the discrete and continuous ICS treatments reaches the maximum when the pulsar evolution time is $\approx50$ kyr and then becomes stable. Consequently, the correction provided by the discrete treatment for this source is expected to be relatively significant. In addition, there are a number of sources in the LHAASO catalog associated with pulsars with an age of $\sim10$~kyr. Among them, the high-energy part of the spectrum of 1LHAASO J1954$+$2836u is relatively hard, which means that if there is a high-energy cutoff in its spectrum, it is more likely to be detected in the future.

We assume that for 1LHAASO J1954$+$2836u, the electron injection spectral index is $\alpha = 2.62$ and the cutoff energy is $E_{e,\mathrm{cut}}=1.17$~PeV. With these parameters, the gamma-ray spectrum derived from the discrete treatment of ICS is consistent with the current LHAASO measurement within the error range. We assume the same $B=1$~$\mu$G for the synchrotron energy loss as in Section~\ref{subsec:geminga}. Additionally, we increase the maximum electron energy for the simulation of energy evolution to $10$~PeV.

First, we obtain the gamma-ray spectrum based on the discrete calculation of ICS. Then, assuming that the LHAASO data size is doubled (with the statistical error reduced by a factor of $\sqrt{2}$), we generate a set of expected spectral points for 1LHAASO J1954$+$2836u. Next, we fit this set of spectral points using the continuous approximation model of ICS. By comparing the difference between the parameter estimation results and the true values, we can determine the systematic bias in parameter estimation that arises from the inaccuracy of the continuous approximation.

\begin{figure*}[htbp]
    \begin{center}
        \subfloat{\includegraphics[width=0.45\textwidth]{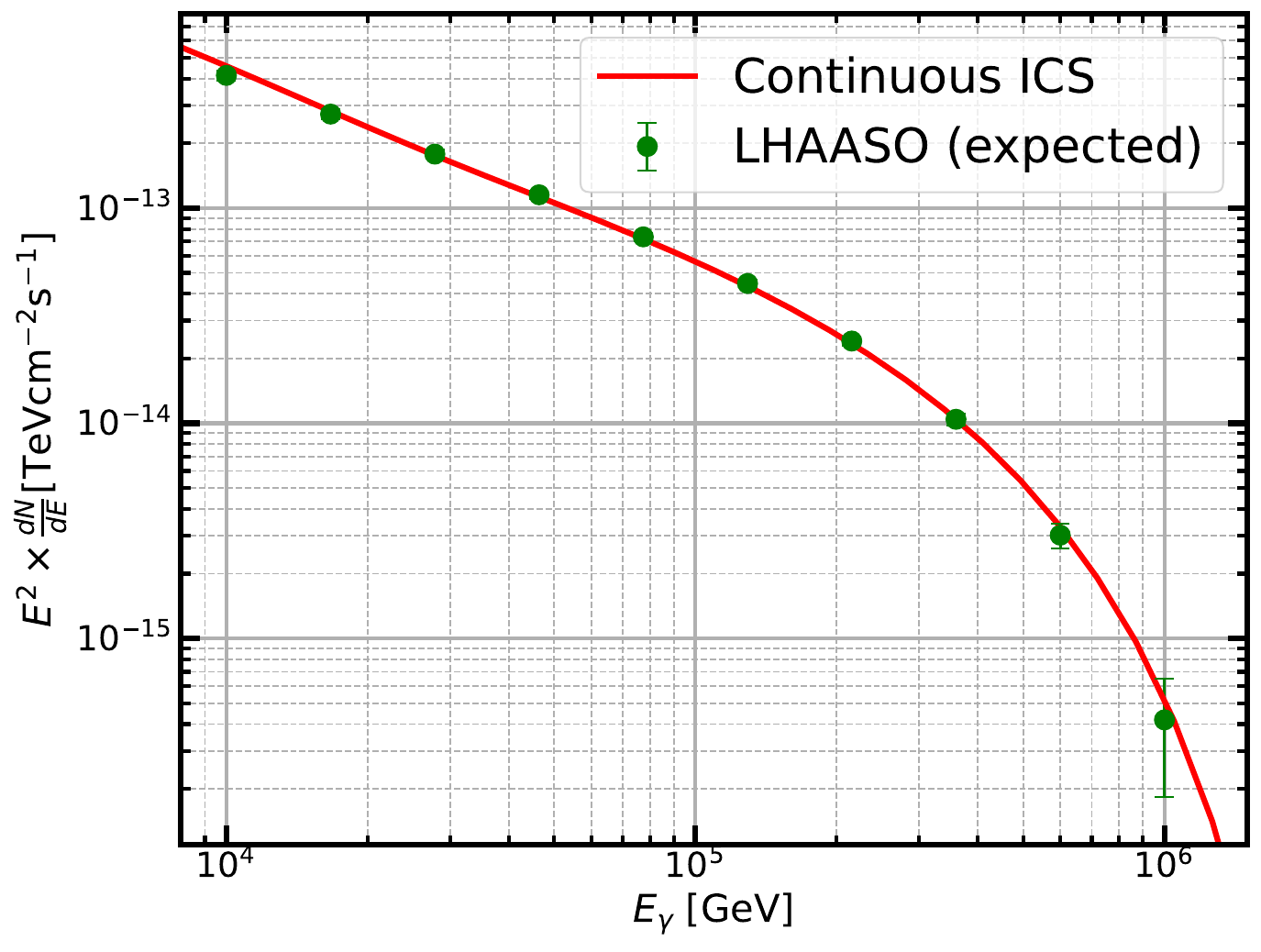}} \hskip 0.01\textwidth
        \subfloat{\includegraphics[width=0.43\textwidth]{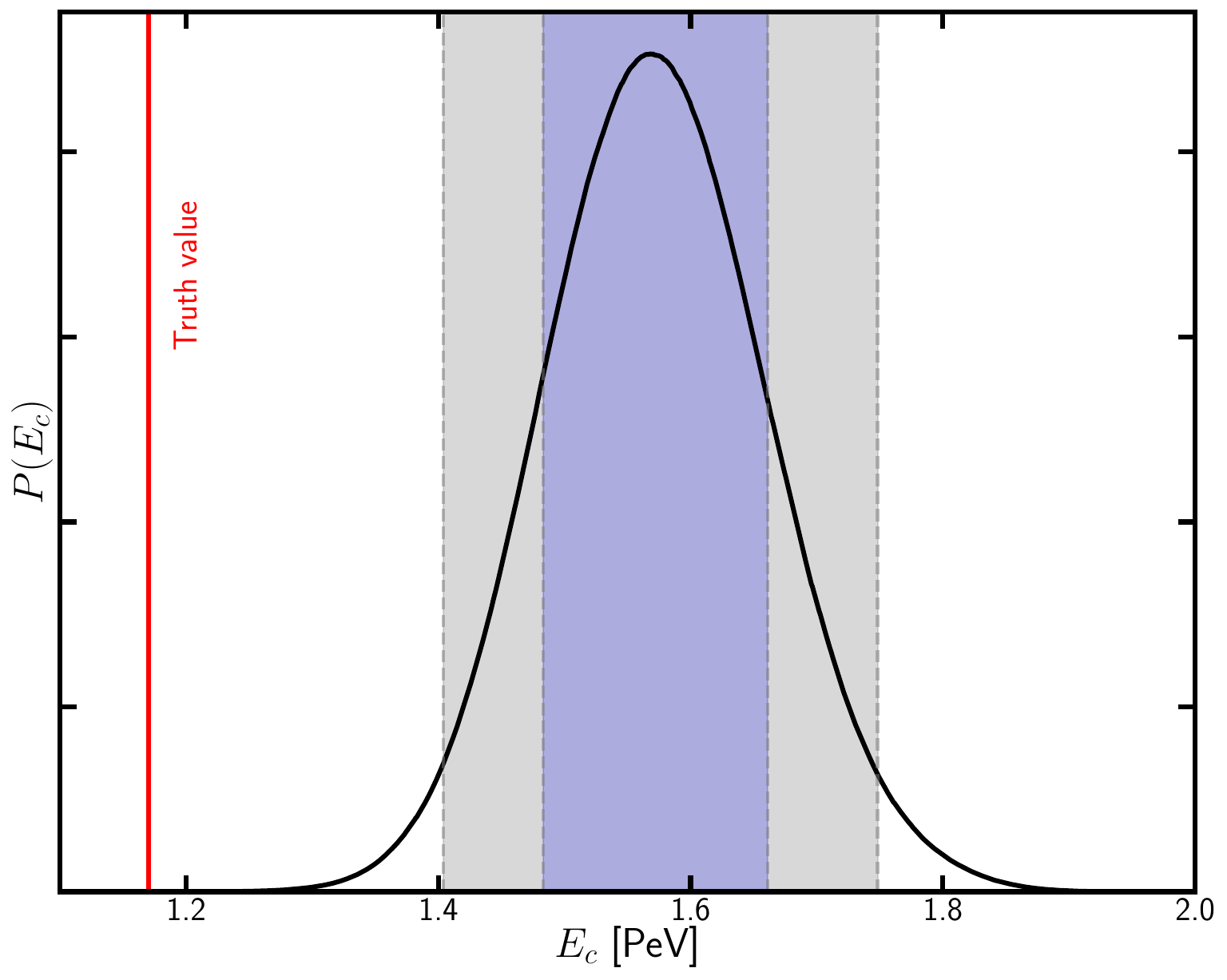}}
    \end{center}
    \caption{Fitting results of the gamma-ray spectrum of 1LHAASO J1954$+$2836u derived from the continuous approximation of ICS to the expected spectral points. The expected spectral points are generated based on the gamma-ray spectrum derived with the discrete treatment of ICS, with the errors determined by assuming the LHAASO data size is doubled. The left panel is the best-fit spectrum compared to the generated data. The right panel exhibits the posterior distribution of the cutoff energy of the electron injection spectrum. The truth value of $E_{e,\mathrm{cut}}$ used in data generation is also shown for comparison, which illustrates that $E_{e,\mathrm{cut}}$ is significantly overestimated if the continuous treatment of ICS is used in calculating the electron energy evolution.}
    \label{fig: lhaaso_1954}
\end{figure*}

Figure~\ref{fig: lhaaso_1954} illustrates the fitting of the gamma-ray spectrum of 1LHAASO J1954$+$2836u obtained from the continuous approximation of ICS to the generated spectral points (left), as well as the posterior distribution of the cutoff energy of the electron injection spectrum determined by the fitting (right). The cutoff energy obtained from the continuous approximation is $1.57^{+0.09}_{-0.09}$~PeV, which is higher than the true value ($1.17$~PeV) with a confidence level of over $4\sigma$. It means that the acceleration capability of the source could be significantly overestimated if continuous approximation is adopted. It should be noted that we use a low magnetic field strength here, so the effect of synchrotron radiation in electron evolution is minor. If the magnetic field strength is higher, the correction amplitude brought by the discreteness of ICS would be reduced, as synchrotron radiation can be safely regarded as a continuous process.


\section{Conclusion}
\label{sec:conclu}
In gamma-ray astronomy, the continuous approximation of ICS is widely used to model the energy evolution of parent electrons responsible for gamma photons. In this study, we use a discrete method to accurately calculate the electron energy evolution caused by ICS. By comparing the results with those obtained from the continuous approximation, we demonstrate the importance of discrete treatment of ICS on parameter estimation in gamma-ray astronomy.

From the perspective of discrete ICS, the initial and final energies of each photon scattered by an electron are both random. Consequently, the energy of evolved electrons is not a deterministic value but rather follows a distribution. We use the Monte Carlo method to simulate the evolution paths of a large number of individual electrons, thereby obtaining the energy distributions of electrons with different initial energies at different evolution times. It should be emphasized that this distribution is physically intrinsic. Provided that the number of electrons simulated is sufficiently large, the resulting electron energy distribution can be considered accurate and reliable.

We apply the results for individual electrons to a time-dependent injection energy spectrum of electrons. The discrete treatment of ICS can lead to significant corrections when the injection spectrum has structures such as a cutoff or a bump. Assuming the electron spectrum has a high-energy cutoff, the cutoff energy of the evolved electron spectrum obtained through discrete ICS treatment is higher than that from the continuous approximation. We find that the difference between the two reaches the maximum when the evolution time reaches $\sim50$~kyr, after which it becomes stable. Furthermore, the correction is more pronounced if the cutoff energy of the injection spectrum is higher.

Pulsar halos and middle-aged PWNe are ideal candidates for applying discrete ICS treatment in gamma-ray astronomy, as they not only have long evolutionary ages but also retain the capability to accelerate electrons to $\gtrsim100$~TeV. Moreover, their magnetic fields are weaker compared to younger PWNe or supernova remnants, which results in a reduced contribution of synchrotron radiation to electron evolution, thereby accentuating the discrete nature of ICS. By fitting the gamma-ray spectrum of the Geminga pulsar halo measured by HAWC, we determine that the cutoff energy of the electron injection spectrum inferred from discrete ICS treatment is $115.0^{+5.4}_{-6.9}$~TeV, whereas the continuous approximation gives $133.7^{+7.1}_{-8.1}$~TeV. This indicates that, at the current precision of HAWC, the continuous approximation systematically overestimates the cutoff energy with a confidence level of $95\%$. We also consider a potential PWN, 1LHAASO J1954$+$2836u, as a case study to evaluate the effect of discrete ICS correction in the PeV regime. Should LHAASO observe the spectral cutoff of this source in the future, the use of continuous approximation may lead to a significant overestimation of the electron acceleration capability. 

As gamma-ray astronomy has now entered an era of ultra-high energy and unprecedented precision, we emphasize the necessity of employing rigorous computational methods, such as the discrete ICS treatment, to ensure that the physical interpretations derived from precise measurements are both accurate and meaningful.

\acknowledgments
We thank Prof. Xiaojun Bi and Dr. Ensheng Chen for their helpful suggestions. This work is supported by the National Natural Science Foundation of China under grants No. 12105292, No. 12375103, and No. 12393853, and the National Key R\&D program of China under the grant 2024YFA1611401.

\bibliography{references}

\end{document}